# Requirements Engineering Current Practice and Capability in Small and Medium Software Development Enterprises in New Zealand

Alison Talbot
Software Engineering Research Lab
Auckland University of Technology
Auckland, New Zealand

Dr Andy Connor
Software Engineering Research Lab
Auckland University of Technology
Auckland, New Zealand
andrew.connor@aut.ac.nz

*Abstract*—This paper presents research on current industry practices with respect to requirements engineering as implemented within software development companies in New Zealand. A survey instrument is designed and deployed. The results are analysed and compared against what is internationally considered "best practice" and previous New Zealand and Australian studies. An attempt is made to assess the requirements engineering capability of New Zealand companies using both formal and informal frameworks

*Keywords-Requirements engineering; Current practice; Capability assessment*

## I. INTRODUCTION

It has been shown that a large proportion of the issues in software development can be related back to requirements engineering (RE) [1]. RE "is the process of discovering the purpose [of the software], by identifying stakeholders and their needs, and documenting these in a form that is amenable to analysis, communication, and subsequent implementation" [2].

Failures during the RE process have a significant negative impact on the overall development process [3]. Reworking requirements failures may take 40% of the total project cost. If the requirements errors are discovered late in the development process, e.g. during maintenance, their correction can cost up to 200 times as much as correcting them during the early stages of the development process [4]. Adequate requirements are therefore essential to ensure that the system the customer expects is produced and that unnecessary efforts are avoided.

From the above it can be seen that missing, poor and changing requirements have all been shown to have a negative impact on project success. It can be argued therefore that good requirements engineering practices that minimise or manage these issues should have a positive influence on project success. As an information economy, New Zealand has a strong history of technology innovation and is currently one of the strongest innovators in the Asia-Pacific region, ranking 5[th] in the regional Deloitte Fast 500 survey [5], with more companies per-capita than any other country in the region. As such, the New Zealand ICT industry provides an opportunity to convey its practices to the rest of the world. However, a number of questions need to be addressed. For example, what constitutes good requirements engineering practice and are these practices actually used in New Zealand? Moreover does the New Zealand software development industry consider such practices relevant, practical or useful?

Davis & Zowghi [6] define a good requirements practice as "a requirements practice that either reduces the cost of the development project or increases the quality of the resulting product when used in specific situations". In this paper no attempt has been made to distinguish between the terms "Good Practice" and "Best Practice" and they are used interchangeably.

Therefore the aim of the research is to determine some of the factors that affect requirements engineering practice particularly within the small and medium sized companies that make up the majority of companies developing software in New Zealand. An additional aim is to discover how New Zealand current practice measures up against both formal and informal process maturity measures. Four research questions are posed:

a) Can capability measures developed overseas be used to assess New Zealand businesses?

b) Have requirements engineering practices in New Zealand changed significantly from those reported in 2000 and 2005?

c) What are the current issues facing RE practice in New Zealand?

d) Are internationally developed "best practices" relevant to New Zealand small and medium sized software development companies?

## II. RELATED WORK

### A. New Zealand Studies

A thorough search has resulted in identifying only four studies of software development practice in New Zealand that have included specific questions about requirements engineering. Only two of these studies [7, 8] focus solely on requirements.

In the first study the researchers created a short list of 65 companies and successfully conducted phone interviews with 24 of these. Following these interviews they undertook in-depth interviews with four of the companies. These were selected from the original 24 in that they represented different sizes and types of applications. Two or three senior staff members in each of the companies were interviewed. These interviews were the basis of their second study.

In the phone interviews the researchers asked questions around the use of tools and formality of the processes involved as well as establishing whether the companies have implemented any industry, New Zealand or international standards. They identify a number of classifications of the type of development that is taking place to which the requirements engineering process relates.

The reported findings from this research are:
- That the size and type of organisation involved in requirements engineering in New Zealand is extremely varied, as are the types of application.
- The larger the software development team, the more likely that well-defined processes are in place. In addition the larger the team, the more time is spent on RE activities and the more rigorous the testing regime.
- Companies involved in developing software for external clients tend to have better defined or formal processes.

From the in-depth interview a number of issues and opinions were identified. These included:
- Commercial considerations sometimes limit the time allowed for capturing requirements.
- Some viewed standards as costly and not delivering competitive advantage, while others held the opposite opinion, once adopted they are easy to maintain and worthwhile
- Software development processes had been in place less than 2 years in many cases. Other noted that documented processes did not reflect the reality.

The third [9] and fourth [10] studies covered the broader topic of software engineering practices in New Zealand. The authors were particularly interested in the adoption of CASE tools, specifically those that support a number of activities in the development process.

The third study involved structured interviews with 5 developers asking about the nature of the project undertaken, experience of the developer, size of company and tools used to support project activities. Findings from this study include:
- Case tools were not being used to their full potential
- Each of the companies had implemented some form of procedures or process for software development
- Less formality was required of developers involved in in-house development as opposed to those developing systems for external clients.

The fourth study used a questionnaire sent to 561 people selected from a business directory that received 147 replies. The questions asked broke the project activities down further than those in the second study. It also focused more on whether the anticipated benefits from the use of CASE tools were being achieved and asked questions about the training received in the use of the tools. Among the findings from this survey were:
- Tools and methodologies are often introduced together
- That use of tools was not mandated
- That integrated tools were not being used mainly because of their complexity and the costs involved
- Many developers placed more importance on processes and methodology than on tools.

The focus of the researchers in all four studies has been on the tools used to support the activities, rather than on any underlying methodologies, although each of them did ask some questions aimed at discovering these.

*B. Australian Studies*

A study of 16 Australian companies covered 28 successful software development projects was carried out in 2005 [11] using both questionnaires and semi-structured interviews. The study looked at the total amount of project effort involved in all RE activities and the proportion of total RE effort that each of the RE activities represented. It also investigated the RE processes followed by the companies and compared these against models from RE literature. Of the 28 projects studied only 5 were developed for external customers. Among the findings were that there was a difference in RE effort between internal and external projects and that more structured processes were evidenced for mission critical and external projects. The study also investigated implicit and explicit RE processes, noting that for external projects the processes were more explicit.

A second study covered in-depth research into the issues relating to multi-site software development in one company [12]. The emphasis in this research was on the Requirements Engineering problems that were experienced and how these were compounded by geographical and cultural differences. The size of the company that was studied is not given, however, given the number and location of sites, it is assumed that the company would be either medium or large, and more likely large. The company was developing a Business Application Environment product that was to be sold and supported from a number of different sites around the world.

Another study [13] concentrated on the relationship between Requirements Engineering and project success. This covered both Australian and US companies. This was a large study involving an initial set of interviews followed by a questionnaire which resulted in 143 responses covering 164 projects, 42 of which came from Australian developers. It covered both internal and external projects and each was classified as either successful or unsuccessful. The majority of the projects were in-house. One of the major findings was that getting good requirements and effectively managing those requirements is a strong predictor of project success. Another finding was that using any software development methodology that includes the RE processes will lead to better results.

*C. Other International Studies*

Many of the international studies involve companies that are much bigger than those that characterise the New Zealand Software development industry [3, 14, 15, 16. 17, 18, 19]. Three studies were found that focused on small and medium enterprises (SMEs).

SMEs in the European context are defined as having less than 250 employees, while small companies are defined as having less than 10 employees. Other criteria such as turnover, are also used in the official European Union definition, however the number of employees appears to be the most widely used criterion [20] and is often the sole criterion for classification.

The first study took place in Germany [21]. It involved a workshop held for 10 companies. This study found that within SMEs
- The maturity level of software engineering is very low
- They typically have "little space for strategic issues such as quality and process improvement"
- There is a large demand for know-how transfer with respect to basic issues
- They are not used to cooperating with external contractors

The companies also noted that the issues they found most relevant to them were modelling, improvement of the requirements document, inspections and tools.

The second study [16, 17] focused on developing a method for RE capability measurement. This model was specifically developed for use in SMEs. Four software development companies were studied, two classified as medium sized (less than 500 employees) and two as small (less than 150 employees). Two companies were located in Ireland and two in Sweden. One of the findings from this study was that the hypothesis that "smaller companies are less likely to have mature RE processes" was not proven. Risk Assessment was identified as a neglected process and the least satisfactory major process area was Requirements Management.

The third study [22] surveyed 12 Finnish companies. Three of these had less than 10 employees, five in the 11-50 range and four over 150. For all companies software development represented the major portion of their business. This survey found that there was:
- A low level of technology transfer in the RE field
- A need to improve their RE practices, requiring development of RE process adaptations, RE process improvement and automation of RE practices.

*D. Summary*

Most of the previous surveys used either interviews or questionnaires or a mixture of both. Both external and internal projects are represented in most of the studies. The focus of the research varies considerably as does the number of respondents. The findings also vary with the findings relating to size of project and formality of RE practices [7] actually the opposite of that found in another study [16, 17]. Some of the findings were common such as those around the effort and degree of formality involved in external projects as opposed to internal projects [7, 9, 11].

III. METHODOLOGY

*A. Survey Design*

Looking at the previous research discussed in section II, most of the studies used either questionnaires, interviews or a mixture of both methods. One used observation [12] and one used a focus group [21]. Bearing in mind the objectives of this research, it was decided to use a questionnaire to reach as many participants as possible, and to follow this up with interviews with selected companies. This paper summarises the survey results only.

The questionnaire contained primarily closed questions that corresponded to those asked in previous surveys, while the in-depth interviews contained a majority of open-ended questions allowing new information to be discovered and further elaboration of the answers to earlier questions. The interviews also provided the opportunity to use the more formal instruments for measuring capability, as these involved asking too many questions for them to be included in the initial survey.

In order to achieve a high return of the questionnaires the number of questions asked was limited to 18 questions. The question design also took into account the amount of explanation needed in order for participants to complete it.

One of the drivers for this research was to build on earlier studies in New Zealand. In Grove, Nickson et al's studies [7, 8], they derived a rough estimate of capability based on formality of processes and formality of language. The questions that allowed this assessment to be made were included in the current research so that comparison could be made. It was decided it was also possible to include the questions on implementation of good practice guidelines relating to the other informal assessment method of RE capability proposed by Nikula, Sajaniemi et al. [22].

Also included in the questionnaire were the demographic questions asked by both the Groves et al study [7] and the Kemp, Phillips et al study [9]. These questions also formed part of the questionnaire used by Nikula, Sanjaniemi et al [22] in their survey on RE Practice. The latter survey was used to source additional questions more specific to requirements engineering than those asked in the New Zealand surveys.

Three new questions were included, aimed at establishing whether Requirements Engineering is carried out by dedicated RE specialists or if it is simply a small part of a software developers duties and to discover whether the people carrying out RE activities had received any formal training in the process. The other new question asked the respondent to identify the top three issues faced by the organisation in relation to software development and requirements engineering. The survey questions and their source are presented in Table I

TABLE I. SOURCE OF SURVEY QUESTIONS

| Source of Survey Question | | | | |
|---|---|---|---|---|
| *Question* | *This Study* | *[7]* | *[9]* | *[22]* |
| 1. What is your Job title? | | | √ | √ |
| 2. Number of employees in New Zealand | | √ | √ | √ |
| 3. Number primarily engaged in software development | | | √ | √ |
| 4. Number primarily engaged in RE | √ | | | |

| Source of Survey Question | | | | |
|---|---|---|---|---|
| Question | This Study | [7] | [9] | [22] |
| 5. Number involved in RE as a minor part of their job | √ | | | |
| 6. Any employees with formal training in RE | | | | |
| 7. Type of Software Development | | √ | | √ |
| 8. Application Types | | √ | √ | √ |
| 9. Typical length of Software Development projects | | | √ | √ |
| 10. Typical no of employees involved | | √ | √ | |
| 11. Requirements Engineering % of total project effort | | √ | | |
| 12. Formality of the requirements engineering process | | √ | | |
| 13. Formality of the language used in requirements specifications document | | √ | | √ |
| 14. Use of tools to support RE processes | | √ | √ | √ |
| 15. Awareness of other tools | | √ | | √ |
| 16. Organisation's implementation of requirements engineering practices. | | | | √ |
| 17. Top 3 issues facing your organisation with respect to software development and requirements engineering | √ | | | |
| 18. Importance of requirements engineering issues to your organisation? | | | | √ |

Only questions 1 and 17 are open-ended questions, while for the remainder of the questions the respondent chose an answer from a number of options presented to them. This was done in order to keep the time required to complete the survey as short as possible, while still providing sufficient richness of data to achieve the purpose of the research.

*B. Selection of Participants*

The New Zealand UBD online business directory was used to source 374 possible participants. This included all companies that were listed under the classification of Software Development. The companies' web sites were visited to discover contact email addresses. As this information was extracted a number of discrepancies were noted between the websites and the UBD directory. There were a significant number of broken links as well. Where the online directory did not provide a usable link a Google search was used to find the websites. Of the 331 companies listed in this category email addresses were initially found for 155 companies.

The initial email invitation was sent to 155 companies in the UBD directory. This resulted in 14 responses. Further investigation using Google and follow up on undeliverable emails resulted in a further 27 emails addresses being discovered. From these companies 3 further responses were received. A search of the Yellow Pages online directory identified a further 53 companies and for those 35 email addresses were found. A second email was then sent to those businesses resulting in 3 more responses, giving a total of 20 responses in all.

As it was felt that this was not a high enough number of responses, it was decided to include the category "Computer Consultancy" from the on-line directories. Surprisingly there was very little overlap with the businesses listed under "Software Development". Using this criterion 338 companies were discovered for which 184 email addresses were found using the same procedures as earlier. The email invitation was sent to these companies resulting in 10 additional responses.

IV. SURVEY SUMMARY

*A. Job Title*

A large majority (86.67%) of the respondents identified themselves as having managerial responsibilities, with only 4 claiming a purely technical role. Only one respondent described themselves as having primarily RE activities giving their title as Business Systems Analyst. The remaining technicians described themselves as software developer or analyst / programmer. This result was unexpected and highlighted the need to be more specific in the way this question is asked. While it is possible that many of those who gave their title as director or similar designation could also be technicians, this cannot be assumed. What it does mean is that the responses represent a management viewpoint rather than that of technicians.

*B. Number of Employees*

Looking at the size of company, 93% (28 companies) would be classified as small (20 or fewer employee) while the remaining 7% (2 companies) classified as medium (21 to 50 employees) using the classification adopted by Phillips et al (2005).

These results are in contrast to other studies which have a much wider spread of company size. However they are consistent with New Zealand Government official company statistics which show that 97% of New Zealand companies have fewer than 20 employees [23].

As mentioned above the Phillips, Kemp et al [10] study targeted software developers rather than companies. It also differed from this study in that it covered all industries, not just software development and computer consultancy companies which represented 47.5% of their respondents. A rough calculation using the midpoint of the range to represent the average number of employees in that group shows the average number of employees in this study is 8.7 compared with 15.3 in the Phillips, Kemp et al [10] study. This size difference may also show in the other questions that relate to number of employees.

*C. Number Primarily Engaged in Software Development*

From the responses, only 3% of companies have greater than 21 employees engaged in software development and 7% of companies have between 11 and 20. The remaining companies are nearly equally divided between having 3-10 employees (47%) and less than 3 employees (43%) primarily engaged in software development. There is a positive correlation between the number of employees and the number engaged primarily in software development. One exception to this trend is respondent number one, who said that less than 3 of their five to twenty employees are involved in systems development while three to five are involved primarily in requirements engineering and 11-20 involved in requirements engineering as a minor part of their job. It appears therefore that this respondent does not consider requirements engineering as part of the systems development process.

### D. Number Primarily Engaged in RE

The survey responses indicated that 67% of the companies have dedicated requirements engineers. As this was a new question, no comparison can be made with other studies, however, given the comparatively small size of the companies, this was an unexpected result. Further analysis of the answers showed that eight out of the ten companies reporting no employees engaged primarily in Requirements Engineering have less than five employees and less than 3 employees engaged in software development. The other two companies both had more than 20 employees more than 10 of whom are primarily engaged in software engineering. One of these companies reported more than 10 employees are involved in RE as a minor part of their job, while the other reported that none were involved in RE as a minor part of their job. This same respondent identified that some of the employees have had formal training in Requirements Engineering, thus indicating an understanding of the term. One possible explanation for this response is that this company is involved only in developments where the Requirements Engineering is undertaken by another party.

### E. Number involved in RE as a minor part of their job

This question aimed at finding the number of employees for whom Requirements Engineering is only a small part of their job. Only six companies (20%) indicated that none of their employees were in this category and four of these companies reported having employees whose primary responsibility is Requirements Engineering. The other two companies are respondent number 28, which is discussed in the previous paragraph and respondent number 15. This latter company has either 3 or 4 employees engaged in software development and reported that none of their employees have any formal training in Requirements Engineering. It is possible that in this case the term requirements engineering may not have been understood by the respondent who identified themselves as the Managing Director. Interestingly this same respondent cited eliciting requirements as their number 1 issue later in the survey. From this and the other answers it seems that this company may also rely on another party to develop requirements specifications for them.

### F. Any Employees with Formal Training in RE

The intention of this question was to discover whether those involved in Requirements Engineering have undertaken any formal training in the process. Ideally those primarily involved in Requirements Engineering would be expected to have some formal training. The wording of this question was slightly confusing as it failed to restrict responses to just those involved in RE. However, the results indicate that 64% of companies have employees with some formal training. Of those that indicated that none of their employees have had formal training (11 companies) only one of these reported as having no involvement in RE activities while five of them have employees whose primary task is RE. This means that in at least 25% of the 20 companies with RE specialists, those specialists are without formal training.

### G. Types of Software Development

The first question aimed at identifying the type of development that was undertaken. The results showed a mix between customised, one-off and package development.

### H. Application Types

This question sought to establish the types of applications being developed. Respondents were asked to rank each of the given application types as being developed never, sometimes, often or always. Respondents could select as many application types as they wished. **Error! Reference source not found.** shows the frequency of development of the different applications, while **Error! Reference source not found.** shows the totals for each application type regardless of frequency. Replies of never or no response have been excluded from these results. From these it can be seen that all application types were well represented within the sample. Comparing results against previous surveys it is noted that there is a close correlation between the applications reported in the Phillips Kemp et al survey [10]. This is shown in Table II.

TABLE II. APPLICATION AREA FOR DEVELOPED SOFTWARE

| Comparison of software development applications | | |
|---|---|---|
| | *This Survey* | *Phillips, Kemp et al [10]* |
| Management Information | 21% | 25% |
| Transaction Processing | 21% | 24% |
| Real Time | 12% | 14% |
| Web | 22% | 21% |
| System | 8% | 8% |
| Embedded | 8% | 8% |
| Other | 8% | |

Other applications that were listed were server technologies, engineering software development, scientific applications or integration of applications, integration to 3D CAD systems and web applications.

### I. Typical Length of Projects

Only 17% of projects are typically longer than 6 months, and 84% were typically between 3 months and a year. Given the generally small size of the companies involved this is not an unexpected result. The Phillips, Kemp & al survey [10] reported that of the projects reported on 60% of the respondents worked on a project lasting between 3 months and one year. However direct comparison may not be valid as in the latter survey respondents were reporting on a single project, while this research asked for the typical length of a project requiring the respondent to consider more than one specific project.

*J. Typical Number of Employees Involved in Projects*

This question, as the previous question did, looked at getting a picture of a typical software development project. Combining the two sets of answers it can be seen that typically the projects are of a short duration (6 months or less) and involve only one or two employees. Using the classification scheme from the Groves, Nickson et al study [8], projects with less than 4 people involved would be small, and those with four to nine people involved medium, then all the projects are either small or medium sized.

*K. Requirements Engineering as a Percentage of Total Project Effort*

This statistic has been included in a number of studies, with their findings showing a wide range of results. The responses in this study also showed a wide variation ranging from 0-5% to 50-55%. The mean is 18%, the standard deviation 12.13% and variance 1.5%. This is a much wider variation than that recorded by MacDonell and Shepperd [24] but is consistent with the findings of Groves, Nickson et al [8]. The mean is close to that reported by Chatzoglou and Macaulay [25], but higher than Groves Nickson et al study [8].

*L. Formality of RE Process*

This is the first of two questions designed to give an informal measurement of the maturity of the RE process of the organisation. The intention of this question is to establish the degree of formality of the RE process in respect of how well it is defined, documented and followed. The responses showed no positive correlation between the number of employees involved in software development and the degree of formality as indicated in **Error! Reference source not found.**. This is in contrast with the Groves, Nickson et al study [8] which found that "it appears that larger software development groups typically have more well-defined software development processes". However it may be that this study had too few larger software development groups for this to be tested.

*M. Formality of Language used in RE Specification Document*

This question extends the previous question by looking at the specification document, the main output from the RE process. The results, shown in **Error! Reference source not found.**, are better than those reported by Neill and Laplante [19] who found that once the no-response figures were removed that 60% of their respondents reported that the language used in specifications was informal, 31.5% semi-formal and 8.5% formal.

*N. Use of Tools to Support the RE Process*

In this part of the survey respondents were asked to state whether they used tools to support the RE process and if so, what tools were used. Of the 28 respondents who answered this question 14 (50%) of them said that no tools were used, while two just specified Microsoft Word. These results are similar to those in Phillips, Kemp et al's [10] study which found that 48.3% of those carrying out requirements gathering used some form of tool to support this activity. The tools used indicate the same range of software development methodologies including structured, prototyping, object-oriented and data-centred identified in the earlier study.

*O. Awareness of Other Tools to Support the RE Process*

The intention with this question was to find the level of awareness of different RE tools. While 11 of the 28 respondents (39%) said that they were aware of other tools only 4 (14%) were able to name specific tools. The proprietary tools named were Rational (2 respondents), Trac, Together and Doors (1 respondent each). Another two respondents listed UML, while one listed imaging software. Although Groves, Nickson et al [8] included this question in their survey, the responses are not included in their analysis, so no comparison can be made with this study.

Of the 14 respondents who are not using any tools, 6 (43%) said that they were aware of other tools but only two (14%) were able to name any. Nikula, Sajaniemi et al's [22] study found that of their 12 companies, none used a commercial RM tool, four (33%) recognised more than one tool, two (17%) recognised one and six (50%) didn't recognise any. It is difficult to make a direct comparison between the studies because the companies in this study were considerably smaller and the respondents were not given a list of tools to select from.

*P. Requirements Engineering Good Practice Guidelines*

In this section of the survey respondents were asked to indicate the level of implementation of the ten guidelines that Sommerville and Sawyer recommend in their book (1997) as the most important RE good practice guidelines. They were given the options of never (scores 0), sometimes (scores 1), normally (scores 2) and compulsory (scores 3). The values were then summed to give a score out of a maximum of 30 points. If a level wasn't attributed to a guideline a "never" response was assumed.

A summary of the scores is shown in **Error! Reference source not found.**. The highest score was 24, the average 10.23, the median 10.5 and the standard deviation 5.37. Comparing these results against Nikula, Sajaniemi et al's [22] findings this study showed more companies in the mid range with a smaller spread of scores (see **Error! Reference source not found.**). This indicates an overall higher level of maturity in the New Zealand companies using this particular measure. It is also worth noting that 5 of the companies (17%) have implemented all ten guidelines at least to some extent. With the highest score being 24 however, using these criteria there is still room for improvement in all the companies.

*Q. Top Three Issues*

Respondents were asked to name the top three issues their organisation faced with respect to software development and requirements engineering. This was an open question aimed at providing the organisations with an opportunity to identify the major issues they faced without being influenced by the researcher.

The answers were grouped into broad categories to allow the responses to be summarised. The classification is subjective in that the discretion of the researcher has been used in interpreting the intent of the respondents. In total, 14 issues were identified with three clearly more common than the others. These three issues were:

1. Scope creep / changing requirements
2. Client acceptance of time/cost/effort spent on requirements before build starts
3. Quality of specification (correctness, clarity, completeness)

While the top issue relating to the management of changing requirements is one that is mentioned frequently in previous studies [1, 22], the second in the list is one that is not asked about in overseas studies. This issue relates to the acceptance by the customer of the time, cost and effort involved in establishing the requirements for the software development.

However this issue is mentioned under unsolicited comments in the Groves, Nickson et al [7] study where some respondents referred to clients were often being impatient to see results and also that commercial considerations prevented them (the companies) from spending more time on capturing requirements, especially where tenders were involved. This impatience is also referred to by Ralph Young in his foreword to Alexander and Stevens book on writing requirements (2002). Other articles discuss the perceived high cost of RE activities [26, 27]. The need to educate the client about the benefits of being thorough in the RE process is also mentioned by Zowghi and Coulin [28].

The individual comments in this classification include "getting people to realise you have to pay for it", "cost to client and perceived benefit", and "convincing customers of its necessity". This indicates that educating the client in the purpose and selling them the benefits of the RE process is a major issue for software development companies in New Zealand. This could be a reflection of the typically small size of the companies surveyed, in that the respondents, mainly directors of the companies, are more likely to be involved in the process of negotiating costs with the customer than would be a software developer or requirements engineer in a large software development company.

In their study Hall et al [3] differentiated between organisational-based and process-based problems. Their three most frequent process based problems were vague initial requirements, undefined requirements process and requirements growth. These items were all of significance to the respondents in this study. In contrast user communication was the fifth most frequent organisational-based problem and the only organisational-based problem they identified that was mentioned in this research.

*R. Importance of Specified Issues to the Organisation*

This question looked at whether some of the issues that had been identified in earlier studies were of any importance to the companies. The most important issue was communication with the customer about requirements, closely followed by the quality of requirements and managing changing requirements. At the other end of the scale, the least important issue was introduction of requirements management tools, which in turn was only slightly less important than the introduction of more formal specification methods. These rankings did not correspond fully with the answers to the previous question. This could be because in this question the respondents did not have to come up with the issues themselves, all they were asked to do was to rate the importance of the issues presented to them. So while the results appear to be contradictive, it does not mean that they are wrong, just that they come from different perspectives.

Comparing the importance of these issues with the Nikula, Sajaniemi et al [22] study, the results are quite different. In that study RE process improvement and RE tool introduction were ranked second and third, while customer communication ranked fifth under general RE development needs. Completeness and change management were the top two issues under requirement development.

V. INFORMAL ASSESSMENT OF RE PROCESS MATURITY

The answers from two questions (12 and 13) in the survey were combined using to produce an informal measure of the RE maturity of the companies based on the degree of formality of the processes followed. Whilst other approaches such as CMMI-DEV could have been used, the focus of this initial survey was on a low overhead analysis based on a small number of questions. This process used the classification in Table III as an aid of estimating maturity.

TABLE III. MAPPING TO INFORMAL RE CAPABILITY MEASURE

| Mapping to informal RE capability measure | | |
|---|---|---|
| **Formality of Process** | **Formality of Language** | **Capability Measure** |
| No explicit process | No formality | No explicit process, No formal language |
| No explicit process | Semi-formal | No explicit process, No formal language |
| Clear phases with informal processes | No formality | Clear phases, no formal language |
| Clear phases with informal processes | Semi-formal | Clear phases, informal specifications |
| Clear phases with a mixture of formal and informal processes | No formality | Clear phases, informal specifications |
| Clear phases with a mixture of formal and informal processes | Semi-formal | Clear phases, informal specifications |
| Clear phases with formal processes | No formality | Clear phases, informal specifications |
| Clear phases with formal processes | Semi-formal | Formal process, semi-formal notation |

This classification is somewhat arbitrary in that the mapping has been specified by the authors based on their understanding of the original questions and how this classification was made in previous studies. Groves, Nickson et al [8] stated in their study that this assessment was "very subjective" because it was reliant on what the interviewer recorded and how this was perceived in the analysis stage of the study. Informal generally implies the use of unstructured natural language, whilst semi-formal includes use of pseudo-code, diagramming techniques (including UML diagrams). Formal languages include Z, B, VDM or other similar approaches though as none of the respondents said that they used a fully formal language for specification, this response was omitted from the mapping. The comparisons of the results from this study are compared with other studies, as shown in Table IV.

TABLE IV. COMPARISON OF INFORMAL RE CAPABILITY RESULTS

| Comparison of Informal RE Capability Results | |
|---|---|
| *Study* | *Findings* |
| Groves, Nickson et al [8] | 17% No explicit process, No formal language<br>29% Clear phases, no formal language<br>25% Clear phases, informal specifications<br>29% Formal process, semi-formal notation |
| Hofmann and Lehner [1] | Only some projects defined their RE process explicitly or tailored an organizational process<br>Most stakeholders perceived RE as an ad hoc process |
| Nikula, Sajananiemi et al [22] | None used formal languages for their specifications, while ten(83%) of the companies reported that semi-formal languages were used either as standard (17%), normally (33%) or discretionary (33%), with only two (17%) saying that they were never used. RE process being defined was standard for four companies (33%), normal for one company (8%), discretionary for three companies (25%) and never for four companies (33%) |
| Sadraei, Aurum et al [11] | In most cases, the companies do not have a standard RE process definition<br>In general, RE is generally performed in a particularly ad hoc manner |
| Verner, Cox et al [13] | Requirements were gathered using a specific method in 53% of successful projects and 50% of unsuccessful projects |
| This study | 26.9% No explicit process, no formal language<br>7.7% Clear phases, no formal language<br>57.7% Clear phases, informal specifications<br>7.7% Formal process, semi-formal notation |

Additionally, while not providing statistics, Hofmann and Lehner [1] in their study of fifteen projects, also noted that "only some projects defined their RE process explicitly or tailored an organizational process" and "most stakeholders perceived RE as an ad hoc process". In their discussion on RE process awareness Sadraei, Aurum et al [11] say that "in most cases, the companies do not have a standard RE process definition", and "in general, RE is generally performed in a particularly ad hoc manner". This is very similar to Hofmann, Lehner et al's findings.

Verner, Cox et al [13] found in their study of RE and software project success covering 164 projects, that requirements were gathered using a specific method in 53% of successful projects and 50% of unsuccessful projects again indicating a relatively low level of RE process maturity using this measure.

Nikula, Sajananiemi et al [22] report that of their twelve companies, none used formal languages for their specifications, while ten of the companies (83%) reported that semi-formal languages were used either as standard (17%), normally (33%) or discretionary (33%), with only two (17%) saying that they were never used. They also reported that the RE process being defined was standard for four companies (33%), normal for one company (8%), discretionary for three companies (25%) and never for four companies (33%). These results show a slightly higher level of maturity overall than the other studies.

Direct comparison is only possible with the Groves, Nickson et al study [8], where the same classification was used. The results from this study show perhaps a slight improvement in the maturity of RE processes in that in the original study 54% of the respondents were classified as either clear phases with informal specifications, or formal process with semi-formal notation, compared to this study where 65.4% were in this category. However, the original study had more companies in the higher classification 29% compared with 7.7%., and fewer at the lowest level 17% compared with 26.9%. This somewhat negates the previous statement. It is also worth considering the difference in the sizes of the companies included in the two studies. It seems therefore that using this measurement, there is no valid conclusion that can be made as to whether the level of maturity of the RE process in New Zealand has changed in the ten years since the original study.

VI. CONCLUSIONS

The aim of the research was to determine some of the factors that affect requirements engineering practice particularly within the small and mediums sized companies that make up the majority of companies developing software in New Zealand and to discover how New Zealand current practice measures up against both formal and informal process maturity measures.

With respect to the first hypothesis "that the capability measures developed overseas can be used effectively to assess New Zealand businesses", insufficient responses were received to allow the testing of formal capability measures. In addition the results from the informal measures were inconclusive.

The second hypothesis was that "requirements engineering practices in New Zealand have changed significantly from those reported in earlier studies". While this study targeted a different population than the earlier New Zealand studies, there appeared to be little difference in the type of software being developed, the methodologies and tools used or the formality of the processes involved. Thus this hypothesis was not proven.

The third hypothesis was that "the issues facing RE practice in New Zealand software development companies are the same as those faced by similar sized overseas companies". This hypothesis has been borne out by the correlation between results from overseas studies and the responses to this survey, both in respect of the rating of the given issues and in the top 3 issues that respondents identified.

The final hypothesis was that "international best practices are not relevant to RE practice in small and medium sized New Zealand software development and computer consultancy companies". The results relating to this hypothesis are inconclusive. Certainly the issues are just as relevant, but the lack of interviews meant that it was not possible to test this.

It is unfortunate that conclusive results were possible for only one of the four hypotheses, namely that the issues faced by New Zealand companies are much the same as those faced by similar sized overseas companies. The first and fourth hypotheses were unable to be tested because companies were not willing to be interviewed. Finally a more representative sample that more closely matched the profiles of the companies involved in the earlier New Zealand Surveys [7, 8, 9, 10] would be needed in order to fully test the hypothesis that the RE practices in New Zealand have significantly changed since those earlier studies. Future research is planned to extend this work and deal more effectively with these limitations.

In summary, this research has looked at requirements engineering practice in the software development industry in New Zealand. It has focused on discovering current practice and made comparisons against previous New Zealand studies and studies overseas that relate to small and medium sized companies. The emphasis has been on attempting to measure the maturity of RE practices and the issues that the industry faces in respect of RE. Whilst further detailed analysis is required, it appears that the issues and challenges faced by New Zealand companies are similar to those faced by similar overseas companies. Yet despite these issues the New Zealand ICT industry continues to perform competitively in the Asia-Pacific region.